\documentclass[aps,prb,preprint,groupedaddress, showpacs]{revtex4}
\usepackage[english]{babel}
\usepackage[latin1]{inputenc}
\usepackage{amssymb,amsmath}
\usepackage{graphicx}
\DeclareMathOperator{\proj}{P}
\DeclareMathOperator{\anti}{A}
\DeclareMathOperator{\sign}{sign}

\begin{document}

\title{Composite fermions from the algebraic point of view}
\author{V. Ruuska}
\author{M. Manninen}
\affiliation{Department of Physics, NanoScience Center,  P.O. Box 35
  (YFL), FIN-40014 University of Jyvaskyla}
\date{\today}

\begin{abstract}
Composite fermion wavefuctions have been used to describe electrons in a strong magnetic field. We show that
the polynomial part of these wavefunctions can be obtained by applying a normal ordered product of suitably defined 
annihilation and creation operators to an even power of the Vandermonde determinant, which can been considered as a
kind of a non-trivial Fermi sea. In the case of the harmonic interaction we solve the system exactly in the lowest Landau
level. The solution makes explicit the boson-fermion correspondence
proposed recently.
\end{abstract}


\maketitle

\section{Introduction} \label{intro}

Composite fermions provide a unified approach to fractional quantum Hall effect (FQHE)~\cite{Tsui}  and other liquid states of 
the two-dimensional electron gas~\cite{Halperin}  in a strong magnetic
field. They have also been used to
describe rotating states of quantum dots~\cite{Jain1}. The composite
fermion wavefunctions proposed by Jain~\cite{Jain2, Jain3}
are obtained in the following manner.
First multiply a wavefunction $\phi$ of {\it non-interacting} electrons by the Jastrow factor
\begin{equation*}
D^k= \prod_{i<j} (z_i-z_j)^{2k}
\end{equation*}
and then project the product to the lowest Landau level. The ansatz was motivated by a heuristic derivation
in which an even number of flux quanta were attached to the electrons and a mean-field approximation was applied to 
the resulting state. Numerical studies for finite systems have supported the ansatz. Recently similar numerical studies 
have provided additional information about quantum dots. In particular, rotating systems of a number of identical particles 
with repulsive interactions confined in a trap have been studied. It has been found that the systems form vortices and that
the general features of the vortex formation are common to both boson
and fermion systems~\cite{Manninen1, Manninen2}.   

In this article we study particles in a harmonic trap using analytical and algebraic means.The physical realization could be
a two-dimensional electron gas in a strong magnetic field or a quantum dot. The mathematical model is basically the same.
In particular, we shall utilize the method of spectrum generating algebras, which is reviewed in section~\ref{ladder}.
Composite fermions are considered in section~\ref{composite}. Our result is a simple  expression
for the composite fermion wavefunctions. The polynomial parts of these wavefunctions are obtained by applying a normal ordered
product of suitably defined annihilation and creation operators to an even power of the Vandermonde determinant, which can been 
considered as a kind of a non-trivial Fermi sea. Finally, in section~\ref{harmonic} we give the  exact solution
for the harmonic interaction in the lowest Landau level. The LLL
spectrum turns out to be qualitatively different from that of the
complete solution, which is also known for the harmonic
interaction~\cite{Poles, Lawson}.  Nevertheless, using the exact eigenstates we show that there is a simple correspondence
between the boson and fermion systems as anticipated by Toreblad et al.~\cite{Manninen1}.

\section{Spectrum generating algebras} \label{ladder}

It often happens in quantum mechanics that energy eigenstates can be obtained from each other by applying suitable
ladder operators. We say that $K$ is a {\it ladder operator} if the commutator of the Hamiltonian $H$ and $K$ is a multiple 
of $K$, that is, $[H,K]=\lambda K$. Then it follows for an energy eigenstate $\phi$ with energy 
$\epsilon$ that $K\phi$ is also an energy eigenstate with energy $\epsilon + \lambda$ since
$$H(K\phi) = K(H\phi)+\lambda K \phi = (\epsilon + \lambda) K\phi.$$
If $K_1$ and $K_2$ are ladder operators such that $[H,K_1]=\lambda_1 K$ and $[H,K_2]=\lambda K_2$, then
\begin{align*}
[H,[K_1,K_2]]&=[[H,K_1],K_2]+[K_1,[H,K_2]] \\
             &=[\lambda_1 K_1, K_2],+ [K_1,\lambda_2 K_2] \\
             &=(\lambda_1+\lambda_2)[K_1,K_2].
\end{align*}
Hence the ladder operators form a Lie algebra $\mathfrak {L} (H)$.
Clearly all the symmetries of the Hamiltonian, that is, the operators commuting with the
Hamiltonian, belong to $\mathfrak {L} (H)$. Loosely speaking, a subalgebra $\mathfrak {g}$ 
of  $\mathfrak {L} (H)$ is called a
{\it spectrum generating algebra} if the physical state space is an irreducible representation of $\mathfrak {g}$.
Instead of making the definition precise, we try to illustrate the concept by the following familiar examples.

The best-known example is the  harmonic oscillator, the Hamiltonian of which is
\begin{equation*}
H= -\frac {1}{2} \nabla^2 + \frac{1}{2} {\mathbf {x}}^2
\end{equation*}
where we have chosen the length and energy units to be $\sqrt {\hbar/m\omega }$  and $\hbar \omega$, respectively. Here $m$ is 
the mass of the particles and $\omega $ is the angular frequence of the oscillator.
In the one-dimensional case all the energy eigenstates can be generated from the unique ground state
\begin{equation*}
\phi_0(x)=\pi^{-1/4}e^{-\frac{x^2}{2}}
\end{equation*}
using the creation operator 
\begin{equation*}
a^{\dagger} =\frac{1}{\sqrt{2}}(\hat x - \frac {i}{\hbar} \hat p)
            =\frac{1}{\sqrt{2}} (x - \frac {d}{dx}).
\end{equation*}
The normalized eigenfunctions are
\begin{equation} \label{onedim}
\phi_n =\frac{1}{\sqrt{n!}}(a^{\dagger})^n\phi_0 
       =\frac{1}{(2^n \sqrt{\pi} n!)^{1/2}}(x - \frac {d}{dx})^n e^{-\frac{x^2}{2}}.
\end{equation}
We have the commutation relations $[H,a^{\dagger}]=a^{\dagger}$ and $[H,a]=-a$ where the annihilation operator
\begin{equation*}
a =\frac{1}{\sqrt{2}} (x + \frac {d}{dx}).
\end{equation*}
In this case the appropriate spectrum generating algebra $\mathfrak {g}$ is the Heisenberg algebra
generated by $a$ and $a^{\dagger}$ with the canonical commutation relation $[a,a^{\dagger}] = 1$.   

\begin{figure}
   \includegraphics{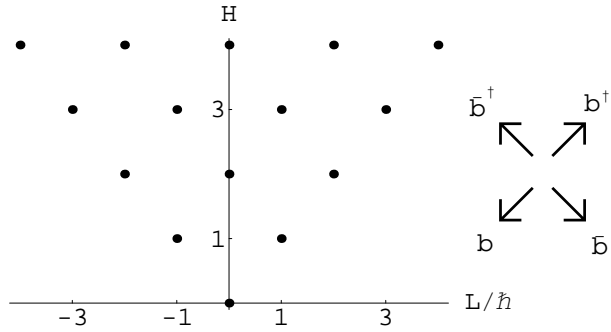}
   \caption{The eigenstates of the two-dimensional harmonic oscillator
     in the (L,H) plane. The action of the annihilation and creation
     operators on the eigenstates is also shown graphically. \label{spectrum}} 
\end{figure}

This approach readily generalizes to higher dimensional oscillators; we only need one pair of creation and annihilation
operators for each degree of freedom. Interestingly, we can do  something slightly different in the plane. Let us use the 
standard notation
\begin{equation*}
\frac{d}{dz}=\frac{1}{2}(\frac{\partial}{\partial x}-i\frac{\partial}{\partial y}), \qquad
\frac{d}{d\bar{z}}=\frac{1}{2}(\frac{\partial}{\partial x}+i\frac{\partial}{\partial y}) 
\end{equation*}
Because of the rotational symmetry the Hamiltonian and the angular momentum operator
\begin{equation} \label{rotation}
L=\hbar(z\frac{d}{dz}-\bar {z} \frac {d}{d \bar{z}})=-i\hbar(x\frac{\partial}{\partial y}-y\frac{\partial}{\partial x})
\end{equation}
commute and we can find a basis of common eigenvectors, which we may express in the following forms:
\begin{equation} \label{twodim}
\begin{split}
\phi_{mn}&=\frac{1}{\sqrt{2\pi m!n!}} \,  e^{\frac{1}{2}\vert z \vert^2} \frac{d^m}{dz^m} \frac{d^n}{d\bar{z}^n} e^{-\vert z \vert^2} \\
          &=\frac{1}{\sqrt{2\pi m!n!}} \,  (e^{\frac{1}{2}\vert z \vert^2 } \frac{d}{dz} e^{-\frac{1}{2}\vert z \vert^2})^m \,
                                             (e^{\frac{1}{2}\vert z \vert^2} \frac{d}{d\bar{z}} e^{-\frac{1}{2}\vert z \vert^2})^n\,
                                             e^{-\frac{1}{2}\vert z \vert^2} \\
          &=\frac{(-1)^{m+n}}{\sqrt{2\pi m!n!}} \,  (\frac{1}{2} \bar {z}-\frac{d}{dz})^m \,(\frac{1}{2} z- \frac{d}{d\bar{z}})^n \,  e^{-\frac{1}{2}\vert z \vert^2}
\end{split}
\end{equation} 
Hence the generators of the spectrum generating algebra can be chosen to be
\begin{align*}
b^{\dagger} &= \frac{1}{2} z - \frac {d}{d\bar {z}}, &
{\bar {b}}^\dagger  &= \frac{1}{2} \bar {z} - \frac {d}{dz }, \\
b &= \frac{1}{2}\bar {z} + \frac {d}{dz}, &
\bar {b}  &= \frac{1}{2} z + \frac {d}{d \bar {z} }
\end{align*}
with the commutation relations $[b,\bar {b}]= [b^{\dagger},{\bar {b} }^{\dagger}] = [b,{\bar {b} }^{\dagger}] =
 [b^{\dagger}, \bar {b}] =0 $ and 
$[b,b^{\dagger} ]= [\bar {b},{ \bar {b} }^{\dagger} ]=1 $. This is again the Heisenberg algebra but the advantage is that
the algebra is spectrum generating not only for $H$ but also for $L$. Using the expression 
\begin{equation*}
H= \frac{1}{2}(b b^{\dagger} + b^{\dagger} b + \bar {b} {\bar {b}}^{\dagger} +{\bar {b}}^{\dagger} \bar {b} )
\end{equation*}
and equation~\eqref{rotation} it is easy to show the necessary commutation relations:
\begin{align*}
&[H,b^{\dagger}] = b^{\dagger}, &[H,{\bar {b}}^{\dagger}] &= {\bar {b}}^{\dagger},
&[H,b] &= - b, &[H,\bar {b}] &=  - \bar {b} \\
&[L,b^{\dagger}] = \hbar b^{\dagger}, &[L,{\bar {b}}^{\dagger}] &= -\hbar {\bar {b}}^{\dagger},
&[L,b] &= - \hbar b, &[L,\bar {b}] &= \hbar \bar {b}.
 \end{align*}
The eigenstates as well as the action of the above generators on them
are shown in Fig.~\ref{spectrum}.

\section{Composite fermions} \label{composite}

Composite fermions~\cite{Jain2, Jain3}  provide a unified approach to
fractional quantum Hall effect (FQHE)~\cite{Tsui}  and other liquid states of the two-dimensional 
electron gas~\cite{Halperin} in a strong magnetic field. They have also been used to describe rotating states of quantum 
dots~\cite{Jain1}. Let us consider electrons confined to the plane and subjected to a magnetic field $H$ perpendicular to
it. Assume that the strong magnetic field makes the electrons effectively spinless. In the symmetric gauge 
\begin{equation*}
\mathbf{A} = \frac{1}{2} H(y\mathbf{e}_x-x\mathbf{e}_y)
\end{equation*}
the single-body Hamiltonian
\begin{equation*}
H_s=\frac{1}{2m}(\frac{\hbar}{i}\mathbf{\nabla}-\frac{e}{c}\mathbf{A})^2
\end{equation*}
takes the form $ H_s = H -\omega  L$ where the angular frequence of the harmonic oscillator is
\begin{equation*} 
\omega = \frac{eH}{2mc}.
\end{equation*}
Note that $\omega$ is half the usual cyclotron frequence but this choice complies with the conventions adopted 
for the harmonic oscillator. Now the common eigenstates $\phi_{m,n}$ of $H$ and $L$ are also the 
eigenstates of $H_s$. However, the eigenenergies change and the ground state becomes infinitely degenerate.
All the states
\begin{equation*}
\phi_{0,n} = \frac{1}{\sqrt{2\pi n!}} \,  z^n \,  e^{-\frac{1}{2}\vert z \vert^2}
\end{equation*}
forming the lowest Landau level (LLL) have zero energy. The Hilbert space
of LLL states may be identified with the Bargmann space 
\begin{equation*}
\mathcal {H}_0= L^2_{hol}(\mathbb {C} \, ; \, e^{-\vert z \vert^2} dm(z))
\end{equation*}
of square-integrable holomorphic functions with respect to the weight function $e^{-\vert z \vert^2}$ while the entire
state space is 
\begin{equation*}
\mathcal {H}= L^2(\mathbb {C} \, ; \, e^{-\vert z \vert^2} dm(z)).
\end{equation*}
Here $dm(z)$ means the Lebesgue measure. Including the exponential factor into the measure reduces the calculations
to polynomial calculus. This technique was already inherent in Laughlin's work~\cite{Laughlin1}, and was explained
by Girvin and Jach~\cite{GirvinJach1}. 

Let $\proj$ stand for the projection $\mathcal{H}  \to  \mathcal{H}_0$. 
Since the functions 
\begin{equation*}
\chi_n(z) =\frac{1}{\sqrt{2\pi n!}} \,  z^n
\end{equation*}
constitute an orthonormal basis of $\mathcal {H}_0 $, the kernel of the projection $\proj$ with respect to the weighted measure
is
\begin{equation*}
K(z,w)=\sum_n \chi_n(z) \overline {\chi_n(w)} =\frac{1}{2\pi} \sum_n \frac {1}{n!} z^n \bar{w}^n = \frac{1}{2\pi} e^{z \bar{w} }.
\end{equation*}

The composite fermion trial wavefunctions are obtained in the following manner.
First multiply a wavefunction $\phi$ of {\it non-interacting} electrons by the Jastrow factor
\begin{equation*}
D^k= \prod_{i<j} (z_i-z_j)^{2k}
\end{equation*}
and then project the product to the LLL. Here $k$ is a positive integer. All indices will run from $1$ to the number $N$ of
the particles unless otherwise stated. Our goal is to make the calculation
of the composite fermion wavefunctions more explicit. The wavefunctions of the non-interacting electrons are 
anti-symmetrized products of the eigenfunctions $\phi_{m,n}$. The projection can be done one complex 
variable at a time. Consider the polynomial parts of the eigenfunctions without the normalizing factor
\begin{equation*}
\varphi_{mn} = \sqrt{2\pi m!n!} \, \phi_{mn} \, e^{\frac{1}{2} \vert z \vert^2}=
 e^{\vert z \vert^2}\frac{d^m}{dz^m} \frac{d^n}{d\bar{z}^n} e^{-\vert z \vert^2}.
\end{equation*}
Then $\varphi_{mn}$ are holomorphic polynomials and we have to calculate the projection $\proj(\psi \varphi_{m,n})$.
Here $\psi$ is the Jastrow factor or more generally an 
arbitrary holomorphic polynomial considered as a function of one variable only.  We have
\begin{equation} \label{first}
\begin{split}
\proj(\psi \varphi_{mn}) &= \int K(z,w) \, \psi(w) \, [e^{\vert w \vert^2}\frac{d^m}{dw^m} \frac{d^n}{d{\bar{w}}^n}
e^{-\vert w \vert^2}] \, e^{-\vert w \vert^2} \, dm(w) \\
&= \frac{1}{2\pi} \int e^{z\bar{w}} \psi(w) \, \frac{d^m}{dw^m} \frac{d^n}{d{\bar{w}}^n} e^{-\vert w \vert^2} \, dm(w) \\
&= -\frac{1}{2\pi} \int \frac{d}{d\bar{w}} [e^{z\bar{w}} \psi(w)] \, \frac{d^m}{dw^m} \frac{d^{n-1}}{d{\bar{w}}^{n-1}} e^{-\vert w \vert^2} \, dm(w) \\
&= -\frac{1}{2\pi} \int z e^{z\bar{w}} \psi(w) \, \frac{d^m}{dw^m} \frac{d^{n-1}}{d{\bar{w}}^{n-1}} e^{-\vert w \vert^2} \, dm(w) \\
&=-z \proj(\psi \varphi_{m,n-1}) = \dots = (-z)^n \proj(\psi \varphi_{m,0}). 
\end{split}
\end{equation}
by induction. On the other hand,
\begin{equation}
\begin{split}
 \label{second}
\proj(\psi \varphi_{m,0}) &= \int K(z,w) \, \psi(w) \, [e^{\vert w \vert^2}\frac{d^m}{dw^m}
e^{-\vert w \vert^2}] \, e^{-\vert w \vert^2} \, dm(w) \\
&= \frac{1}{2\pi} \int e^{z\bar{w}} \psi(w) \, \frac{d^m}{dw^m}  e^{-\vert w \vert^2} \, dm(w) \\
&= \frac{1}{2\pi} \int e^{z\bar{w}} \psi(w) \, (-\bar{w})^m \, e^{-\vert w \vert^2} \, dm(w) \\
&= (-\frac{d}{dz})^m \, [\frac{1}{2\pi} \int  e^{z\bar{w}} \psi(w) \,  e^{-\vert w \vert^2} \, dm(w)] \\
&= (-\frac{d}{dz})^m \, [\proj(\psi \varphi_{0,0})]= (-\frac{d}{dz})^m \, \psi.
\end{split}
\end{equation}
Combining the equations~\eqref{first} and~\eqref{second} we obtain
\begin{equation*}
\proj(\psi \varphi_{mn}) = (-1)^{n+m} z^n \frac{d^m}{dz^m} \, \psi.
\end{equation*}
In general, an N-particle state of non-interacting electrons is an antisymmetrized product 
$\phi= \anti [\phi_{\alpha_1\beta_1} \cdots \phi_{\alpha_N\beta_N}]$ of the single-particle 
wave-functions. Here the antisymmetrization operator is defined as
\begin{equation*}
\anti[\Phi](z_1,\dots,z_N)=\sum_{\sigma\in S_n} (-1)^{\sign (\sigma)} \Phi(z_{\sigma(1)},\dots,z_{\sigma(N)})
\end{equation*}
where the summation is over all the permutations of the indices $1,\dots,N$.
The above calculation easily generalizes to several variables. We can express the
result using the multi-index notation
\begin{equation*}
z^{\alpha} = z_1^{\alpha_1} \cdots z_N^{\alpha_N} \quad \text{ and } \quad 
\frac{\partial^{\beta}}{\partial z^{\beta}} = \frac {\partial^{\vert \beta \vert}}
{\partial z_1^{\beta_1} \cdots \partial z_N^{\beta_N} } 
\end{equation*}
where $\vert \beta \vert = \beta_1 + \cdots + \beta_N$.
The composite fermion wavefunction corresponding to $\phi$ and Jastrow
factor $D^k$ is given up to normalization by
\begin{equation} \label{cpfunction}
\phi^{CF}_{\alpha, \beta,k} = \anti [ z^{\alpha} (\frac {d^{\beta}}{dz^{\beta}} \, D^k) ] e^{-\frac {1}{2} \vert z \vert^2}.
\end{equation}
The formula is analogous to the equations~\eqref{onedim} and~\eqref{twodim} in the sense that all the states can be
created by applying the multiplication and partial derivative operators to the given ground states, Jastrow
factors.  Consequently, it seems natural to search for spectrum generating algebras even for
interacting fermions. However, to proceed further we have to choose the actual form of the interaction.

\section{Exact results for the harmonic interaction} \label{harmonic}

It was already known for the nuclear physicists of the fifties that the harmonic interaction potential
\begin{equation} \label{harmint}
V = \lambda \sum_{i<j} \vert z_i - z_j \vert^2
\end{equation}
in a harmonic trap is exactly diagonalizable~\cite{Lawson}.
The model has also been applied to quantum dots~\cite{johnson}.
A complete set of eigenfunctions is difficult to write down due to the
degeneracy problems but the eigenenergies
and their degeneracies are known~\cite{Poles}. The general solution
has no restrictions with respect to the angular momentum,
whereas the composite fermion wavefunctions are by definition
restricted to the lowest Landau level. The LLL approximation is usually justified
by physical considerations related either to the strong magnetic field
or to the rotational state. We shall find the exact
solutions of the projected Hamiltonian in the lowest Landau
level. Our result generalizes the observation by Girvin and
Jach~\cite{GirvinJach1, GirvinJach2}, who noticed that
Laughlin's functions
\begin{equation*}
\Psi_m = \prod_{i < j} (z_i -z_j)^m e^{-\frac{1}{2}\vert z \vert^2} 
\end{equation*}
for an odd integer $m$ are exact energy eigenstates for $V$. Comparison with the complete solution will hopefully shed 
light on the subtleties of the LLL approximation. At least the LLL solution makes explicit the correspondence between 
the boson and fermion systems suggested in~\cite{Manninen1}. 

To start with, a technical remark is in order. Note that $\lambda$
must be chosen negative to make the interaction repulsive. Moreover, the
absolute value of $\lambda$ must be small enough to make the energy
spectrum bounded from below. For the LLL Hamiltonian even this is not
sufficient. However, if we fix the total angular momentum, a (degenerate) ground state can be found.
The projection $\hat V=\proj V \proj $ of the interaction~\eqref{harmint} to the Bargmann space $\mathcal {H}_0$ was
calculated by Girvin and Jach~\cite{GirvinJach2}. It is given by the formula
\begin{equation*}
\begin{split}
\hat V &= \lambda \sum_{i,j} (\frac {\partial}{\partial z_i}- \frac {\partial}{\partial z_j})(z_i - z_j) \\
       &= N(N-1)\lambda + \lambda \sum_{i,j} (z_i - z_j)(\frac {\partial}{\partial z_i}- \frac {\partial}{\partial z_j})
\end{split}
\end{equation*} 
Multiplication operator $z_k$ is not quite a ladder operator for $\hat V$ since
\begin{equation*} 
\begin{split}
[\hat V, z_k] &= \lambda \, [\sum_{i<j} (z_i - z_j)(\frac {\partial}{\partial z_i}- \frac {\partial}{\partial z_j}), z_k] \\
              &= \lambda \, \sum_{i<j} (z_i - z_j)(\delta_{ik} - \delta_{jk}) \\
              &= N\lambda \,  z_k - \sum_i z_i.
\end{split}
\end{equation*} 
However, it follows from the above calculation that
\begin{equation*}
[\hat V, z_k-z_l] = N \lambda (z_k-z_l)
\end{equation*}
for all $k$ and $l$. Hence multiplication of an eigestate by the linear factors $z_k - z_l$ creates new eigestates
with eigenvalues increased by $N \lambda$. We also have
\begin{equation*}
[\hat V, \sum_k z_k] = 0
\end{equation*}
so that the operator $\sum_k z_k$ creates states which are
denegenerate with the original one. We can 
start with the constant function $\varphi_B = 1$,  which is an eigenstate for $\hat V$ with eigenvalue 
$N(N-1)\lambda $, and multiply it by the above linear factors to obtain new eigenstates. However,
in order to qualify as proper
physical states, they also have to obey the usual permutation symmetries. Let us consider the
bosonic case first. The question is which symmetric polynomials can be represented as linear combinations of the
products of the factors  $e_{kl} = z_k-z_l$. Since the factors $e_{kl}$ span the space 
$C= \{ z_1 + \dots + z_N=0 \}$, which is isomorphic to 
the Cartan subalgebra of the Lie algebra $\mathfrak {su} (N)$ as a
representation of the symmetric group $S_n$,
the problem is solved by the classical invariant 
theory~\cite{Bourbaki}. As long as we are restricted to $C$, the subalgebra of symmetric polynomials obtained from $e_{kl}$ is freely generated by the power sums
\begin{equation*}
q_k = z_1^k + \cdots + z_N^k
\end{equation*}  
for $1< k \leq N$. Now $q_k$ is obtained by restricting a linear combination of
the products of $e_{kl}$  from ${\mathbb {C}}^N$ to $C$. The polynomials must be invariant in the
translations parallel to $(1,1,\dots,1)$ since the factors $e_{kl}$
are. Hence the only possible extension of $q_k$ to ${\mathbb {C}}^N$
is
\begin{equation*}
p_k = (z_1-\bar z)^k + \cdots + (z_N - \bar z)^k
\end{equation*} 
where $\bar z = (z_1+\dots+z_N)/N$ is the center-of-mass. If we
allow $p_1=z_1+\dots+z_N$ among the generators, then by dimensional
counting we obtain a complete basis of the symmetric polynomials. 
The difference is that the factor $p_1$ does not have effect on the eigenvalue while the other power sums
$p_k$ increase it by $Nk\lambda$. Including the Gaussian factors we
obtain the boson eigenstates
\begin{equation*}
\psi_\alpha^B = \prod_{k=1}^N p_k^{\alpha_k} e^{-\frac{1}{2} \vert z \vert^2} 
\end{equation*}
where $\alpha_k$ are non-negative integers. The corresponding eigenenergies are 
\begin{equation*}
E_\alpha^B= N(N-1)\lambda + N\lambda\sum_{k=2}^N k\alpha_k. 
\end{equation*}
Disbelievers can check the result by a straight-forward
calculation. Once the answer is known, it is not that hard since $\hat
V$ is a derivation up to an additive constant. 
The fermion states can be obtained by exactly the same reasoning starting from the ground state
\begin{equation*}
\varphi_F = \prod_{i<j} (z_i - z_j)
\end{equation*}
with the eigenvalue $N(N-1)(1+N/2) \lambda$. The fermion eigenstates are  
\begin{equation*}
\psi_\alpha^F = \prod_{k=1}^N p_k^{\alpha_k} \prod_{i<j} (z_i - z_j) e^{-\frac{1}{2}\vert z \vert^2} 
\end{equation*}
with eigenenergies
\begin{equation*}
E_\alpha^B= N(N-1)(1+N/2)\lambda + N\lambda\sum_{k=2}^N k\alpha_k. 
\end{equation*}
Observe that the fermion states differ from the boson states only by the factor $\varphi_F$, while 
the corresponding eigenenergies are shifted by a constant depending only on the number of particles and the 
strength $\lambda $ of the interaction. Practically all states are degenerate due to the huge symmetry 
of the interaction potential. 

\section{Conclusions}

We have given a short review on the spectrum generating algebras 
and applied them to derive the composite fermion wavefunctions.
As an example we studied the exactly solvable model of harmonic 
interparticle interactions and derived the wavefunctions and 
energy spectra for spinless fermions and bosons. 
In this case there is a one-to-one correspondense between
the fermionic and bosonic wavefunctions: Any fermion wavefunction
is obtained from the corresponding boson wavefunction
by multiplying it with a simple Slater determinant.

\acknowledgments{This work was supported by the 
Academy of Finland and the European Community project ULTRA-1D 
(NMP4-CT-2003-505457).}

\end{document}